\documentclass[prl,twocolumn,aps,showpacs]{revtex4}

\usepackage{graphicx}
\usepackage{color}

\begin{document}

\title{Correlated Topological Insulators with Mixed Valence}
\author{Feng Lu$^*$, JianZhou Zhao\footnote{these authors contributed equally to the paper}, Hongming Weng, Zhong Fang and Xi Dai}
\affiliation{ Beijing National Laboratory for Condensed Matter Physics, 
              and Institute of Physics, 
              Chinese Academy of Sciences, 
              Beijing 100190, 
              China }
\date{\today}

\begin{abstract}
  We propose the local density approximation (LDA)+Gutzwiller method incorporating Green's function scheme to study the topological physics of correlated materials from the first-principles. Applying this method to typical mixed valence materials SmB$_6$, we found its non-trivial $Z_2$ topology, indicating that SmB$_6$ is a strongly correlated topological insulator (TI). The unique feature of this compound is that its surface states contain three Dirac cones in contrast to most known TIs.
 
\end{abstract}

%\pacs{74.70.-b, 74.25.Jb, 74.25.Ha, 71.20.-b}
\maketitle

%
%
% Non-Interacting and interacting TI

Most of $Z_2$ topological insulators (TI)\cite{Fu:2007ei,Fu:2007io,Hasan:2010vc,Qi:2011wt,Bernevig:2006ij} discovered up to now are semiconductors which are free of strong correlation effects and their topological nature can thus be predicted quite reliably by first principle calculations based on density functional theory (DFT)\cite{Zhang:2009ks}. The $Z_2$ classification of band insulators has been generalized to interacting systems by looking at its response to external electric magnetic field, namely the topological magneto-electric effect (TME)\cite{Qi:2009vo}.  A correlated insulator is a TI if the $\theta$-angle defines in TME is $\pi$. Given the TME theory and several simplifications\cite{Wang:2011jm}, however, its application to realistic materials is still absent due to: (1) the lack of suitable compounds; (2) the difficulty to compute reliably correlated electronic structures from the first-principles. In this letter, we study a special class of materials, the mixed valence (MV) compounds, which contain rare-earth elements with {\it non-integer chemical valence}. By combining the Gutzwiller variational approach from the first-principles and the Green's function method for the TME, we found SmB$_6$ is a 3D correlated TI. Interestingly, it has three Dirac cones on the surface, in contrast to most of the known TIs.

% Introduction of MV compounds

Classic MV compounds, such as SmB$_6$ and YbB$_{12}$\cite{Varma:1976}, share the following common features. (i) The x-ray photoelectron spectroscopy (XPS) and X-ray absorption spectra (XAS) contain peaks from both divalent and trivalent multiplets with comparable spectral weight\cite{Allen:1980gi,Chazalviel:1976wb,Beaurepaire:1990}, indicating the valence of Sm or Yb to be close to 2.5.  (ii) A small semiconducting gap opens at least in part of the Brillouin zone (BZ) at low temperature\cite{Cooley:1995jc}.
Previous electronic structure studies indicate that the electrons transfer from Sm (or Yb) $4f$ orbitals to $5d$ orbitals, which leads to fractional occupation in $4f$ orbitals\cite{Antonov:2002tf,Ylvisaker:2009}. From the band structure point of view, the electron-transfer between $4f$ and $5d$ orbitals indicates possible band inversion between them, which is a crucial ingredient to realize TI (if it happens odd number of times in the entire BZ). At the mean while, SmB$_6$ has been recently suggested to be among the possible candidates of the interesting topological Kondo insulators~\cite{Dzero:2010dj,Dzero:2012kx,Miyazaki:2012iq}.
%where the valence of Sm is more close to 3$^+$ with non-zero local moments being screened by the %$5d$ conduction bands.  
Although the analytical studies of MV compounds have been conducted by several groups, the reliable first principle studies, which is crucial to identify the possible topological phases, is still lacking, due to the strong correlation nature of these materials\cite{Olle:2004,Thunstrom:2009}.  Here we report that the local density approximation (LDA)+Gutzwiller method, a newly developed first-principles tool for correlated electron systems, enables us to search for the topological phases in correlated matters. By taking the MV compound SmB$_6$ as an example, we will focus on two key issues: (i) how to compute $Z_2$ topological index with LDA+Gutzwiller; (ii) what's the topological nature of SmB$_6$ with the strong correlation effects among the $f$ electrons?

\begin{figure} 
  \includegraphics[width=0.25\textwidth,angle=90,clip,trim=0 0 0 0]{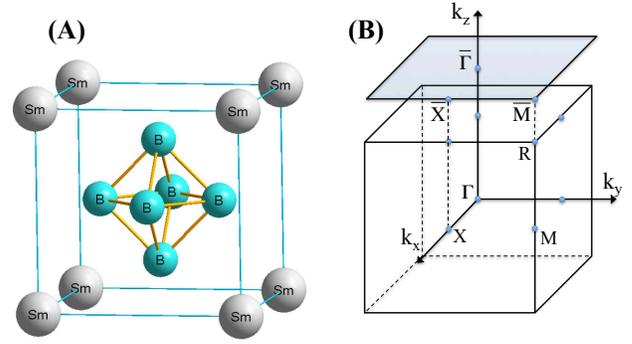} \caption{(A) The CsCl-type structure of SmB$_6$ with $Pm3m$ space group. Sm ions and B$_6$ octahedron are located at the conner and center of the cubic lattice respectively. (B) The bulk and surface BZ for SmB$_6$.}
\label{struct} 
\end{figure}

The LDA+Gutzwiller method combines the DFT within LDA with the Gutzwiller type trial wave function, which takes care of the strong atomic features in the ground state. Here we just sketch the most important aspects, and leaving details to reference~\cite{Deng:2009p6949,RTGW:2012,Yao:2011el}. We start from the common Hamiltonian (used for most of the LDA++ schemes),
\begin{equation}
H_{Total}=H_{LDA}+H_{int}+H_{DC}
\end{equation}
where $H_{LDA}$ is the single particle Hamiltonian obtained by LDA and $H_{int}$ is the local interaction term for the $4f$ electrons described by the Slater integrals $F_0$, $F_2$, $F_4$, $F_6$. In the present study, for Sm atom
 we choose $F_0$=5.8 eV and $F_2=9.89eV$, $F_4=7.08eV$,$F_6=4.99eV$ 
to be their atomic values\cite{Cowan:1981,Carnall:1989}.  $H_{DC}$ is the double counting term representing the interaction energy already considered at the LDA level. In the present paper, we 
compute the double counting energy using the scheme described in reference\cite{Anisimov:1997}.
%{\it (What's the expression for $H_{int}$, what's the parameters F2-F6)}. 
We then use the following Gutzwiller trial wave function, $|G\rangle=P_G|0\rangle=\prod_{i}P_i|0\rangle$, where $|0\rangle$ is a non-interacting state (obtained from LDA), $P_i=\sum_{\Gamma_i}\lambda_{\Gamma_i}|\Gamma_i\rangle\langle\Gamma_i|$ is the Gutzwiller projector at the $i$-th site with $|\Gamma_i\rangle$ being the atomic eigenstates and $\lambda_{\Gamma}$ being the variational parameters to be determined by minimizing the ground state total energy (under Gutzwiller approximation)\cite{RTGW:1998,RTGW:2012,RTSB:2007}, $E_G$=$\langle 0|H_{eff}|0\rangle+\sum_{\Gamma}\lambda_{\Gamma}E_{\Gamma}$. Here $H_{eff}=P_GH_{LDA}P_G$ is called renormalized effective single particle Hamiltonian for the quasi-particles and $E_{\Gamma}$ is the eigen energy of the $\Gamma$-th atomic eigenstate.  The scheme preserves the nice aspect of being variational, however, it is beyond LDA (and also LDA+$U$) because the total energy $E_G$ now relies on the balance between the renormalized kinetic energy of quasi-particle motion and the local interaction energy, which is configuration-dependent. We note 
that including the complete form of all the Slater integrals ($F_0$ to $F_6$) in the local interaction is essential to obtain the correct electronic structure for these materials.

%The the present study, we keep all the atomic eigen states with f electrons to be $4,5$ and $6$, which %totally contains $8100$ different atomic states.

% How to determine topological index by Gutzwiller method

The linear response theory for the coefficient of TME has been developed and simplified by Z. Wang {\it et al}\cite{Wang:2011jm}.  For interacting systems, when the self-energy contains no singularity along the imaginary axis, the formula for the TME coefficient only requires the single particle Green's function at zero frequency, $\hat g(k,0)$. Because the singular point for the self-energy along the imaginary axis only appears for a Mott insulator with completely localized $f$-orbitals, which is not the case for SmB$_6$, we can safely apply the above method here.  Therefore the way to determine the $Z_2$ invariance is just diagonalizing the Hermite matrix $-\hat g(k,0)^{-1}=\hat H_0-\mu_f+\hat \Sigma(0)$ and treating the eigenstates with the negative eigenvalues as the ``occupied states''. If the system also has spacial inversion center, the $Z_2$ invariance can be simply determined by counting the parities of these ``occupied state'' on the time reversal invariant momenta (TRIM) points, the same as that done for the non-interacting topological band insulators\cite{Fu:2007ei,Fu:2007io}.

In the Gutzwiller approximation (or the equivalent slave boson mean field approach), the low energy single particle Green's function of an interacting system can be expressed by the quasi-particle effective Hamiltonian and the quasi-particle weight $\hat z$ (which in general is a matrix) as, 
\begin{equation}
  \hat g(k,i\omega)={{\hat z} \over {i\omega-\hat H_{eff}+\mu_f}}+\hat g_{ic}(k,i\omega) \label{green_gutz} 
\end{equation}
where the second term describes the incoherent part of the Green's function, which is ignorable for low frequency. Therefore by applying the green's function method described above, we reach the conclusion that the $Z_2$ invariance in LDA+Gutzwiller method is determined by the occupied eigenstates of the Gutzwiller effective Hamiltonian $H_{eff}$, which can be interpreted as the band structure of the ``quasi-particles''.

% describe the structure of the material
We now focus on SmB$_6$, which is crystallized in the CsCl-type structure with Sm ions and $B_6$ clusters being located at the conner and body center of the cubic lattice respectively (see Fig.1).  The LDA part of the calculations has been done by full potential linearized augmented plane wave (LAPW) method implemented in the WIEN2k package\cite{Blaha:2001vw}. BZ integration was performed on a regular mesh of 12$\times$12$\times$12 $k$ points. The muffin-tin radii($R_{MT}$) of 2.50 and 1.65 bohr were chosen for Sm and B atoms, respectively. The largest plane-wave vector $K_{max}$ was give by $R_{MT}K_{max} = 8.5$. The spin-orbit coupling (SOC) is included in all calculations.

%The band structure for the bulk materials: LDA results
\begin{figure}
\includegraphics[width=0.42\textwidth,clip,angle=270,trim=0 20 0 0]{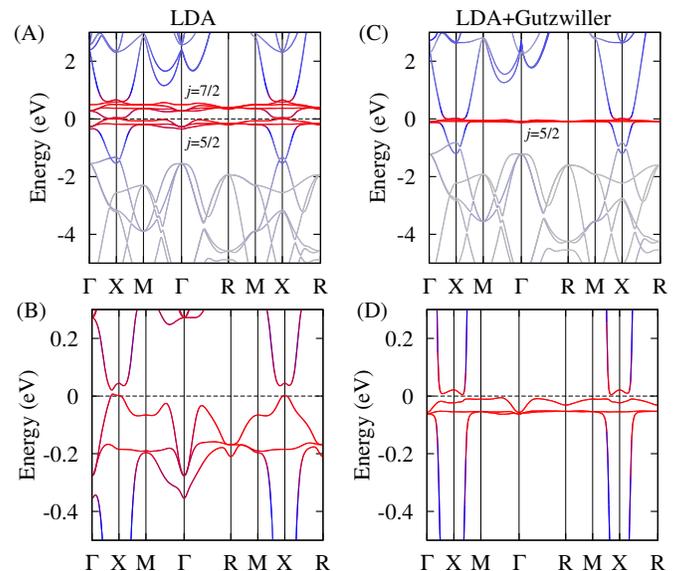}
\caption{The calculated band structure for SmB$_6$ in different energy scales by LDA (A, B) and LDA+Gutzwiller(C, D).The B and D are just the zoom in of A and B respectively, around the fermi level. Compared to the LDA results, the $4f$ $j$=7/2 bands in LDA+Gutzwiller have been pushed up to be about 4.0 eV above the fermi level, leaving the band structure near the fermi level being dominated by $4f$ $j$=5/2 and $5d$ states. 
The blue, red and grey colors represent the weight of the orbital character for $5d$, $4f$ and $2p$ respectively.}\label{band}
\end{figure}

The band structure obtained by LDA is shown in Fig.\ref{band} (A and B), where we can find three major features. (i) The Sm-$4f$ orbitals, which split into the $j$=5/2 and $j$=7/2 manifolds due to the SOC, form narrow bands respectively with width around $0.5$ eV near the Fermi level. (ii) The low energy band structure is semiconductor-like with a minimum gap about 15meV along the $\Gamma$-$X$ direction.  (iii) There is clear band inversion features at the X-points, where one $5d$ band goes below the $f$ bands, which reduces the occupation number of the $f$-states to be around $5.5$.

The first two features of the LDA band structures are not consistent with the experimental observations.  Firstly, the XPS measurements find quite strong $4f$ multiplet peaks, indicating strong atomic nature of $4f$ electrons in SmB$_6$ (in other words, most of the $4f$ electrons are not envolved in the formation of energy bands)\cite{Allen:1980gi,Chazalviel:1976wb}.
%{\color{red} (Please check these citation.)} 
Secondly, both transport~\cite{Menth:1969vq,Nickerson:1971up,Cooley:1995we} and optical~\cite{molnar:1982, Nanba1993440, 
Travaglini:1984wh,Demsar:2006} measurements reveal the formation of a small gap only for temperature below 50 K.
%{\color{red} (Do we need so many references here?)} 
The above two features imply that the $f$ electrons in SmB$_6$ have both localized and itinerant natures and the correct description of its electronic structure should include both of them. The qualitative physical picture can be ascribed to Kondo physics, which involves both itinerant and localized electronic states\cite{Dzero:2012kx}.  At high temperature, these localized orbitals ($4f$ states) are completely decoupled from the itinerant energy bands, forming atomic multiplet states. While in the low temperature, the coherent hybridization between localized and itinerant states is gradually developed leading to the formation of ``heavy quasi-particle'' bands. The insulating behavior appears when the chemical potential falls into the ``hybridization gap'' between the heavy quasi-particle and the conduction bands, which is mainly of the $5d$ orbital character in SmB$_6$.

%The band structure for the bulk materials: LDA +G results

\begin{figure}
\includegraphics[width=0.35\textwidth,clip,angle=270,trim=0 50 0 0]{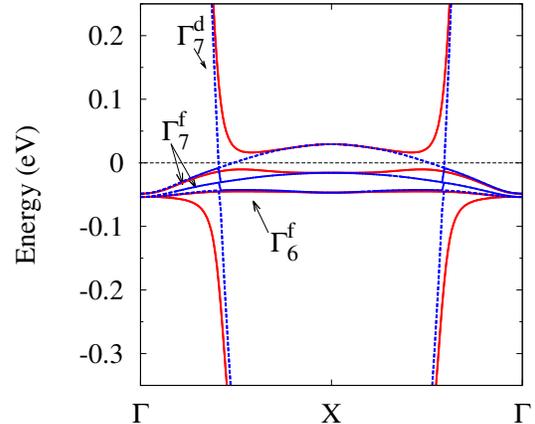}
\caption{Band hybridizations between $4f$ and $5d$ along $\Gamma$ to $X$ direction. The original $4f$ $j$=5/2 orbitals split into two $\Gamma^f_7$ and one $\Gamma^f_6$ bands.  The $\Gamma^f_7$ bands cross and hybridize with another $\Gamma^d_7$ band formed by $5d$ orbitals.}\label{hybrid}
\end{figure}

Such Kondo picture can be nicely captured by our  LDA+Gutzwiller calculations, which provide equal-footing descriptions of both the itinerant $5d$ bands and those heavy quasi-particle states formed by $4f$ orbitals.  The quasi-particle band structures obtained by LDA+Gutzwiller is shown in Fig.\ref{band} (C and D). Compared with the LDA results, there are three major differences induced by the strong correlation effects. (i) The $4f$ $j$=7/2 bands are pushed up to be around 4.0 eV above the fermi level, leaving the band structure near the fermi level being dominated by $4f$ $j$=5/2 and $5d$ bands. (ii) The band width of those ``heavy quasi-particle'' bands formed by $4f$ orbitals are much reduced to be less than 0.1 eV. The quasi-particle weight $z$ of these ``heavy quasi-particle'' bands is about 0.18, indicating that the quasi-particles are formed by less than $20\%$ of the $4f$ spectral weight and the remaining weight is attributed to the atomic multiplets (or the Hubbard bands). (iii) A hybridization gap between $4f$ quasi-particle bands and the itinerant $5d$ bands appears along the $\Gamma$ to $X$ direction. The detail hybridization process is illustrated in Fig.~\ref{hybrid}. Along the $\Gamma$ to $X$ direction, the point group symmetry is lowered to be $C_{4v}$, whose double group contains two irreducible two-dimensional representations, $\Gamma_6$ and $\Gamma_7$. Along $\Gamma$ to $X$, the original $4f$ $j$=5/2 orbitals split into two $\Gamma_7$ and one $\Gamma_6$ bands, which cross with another $\Gamma_7$ band formed by $5d$ orbitals. The hybridization terms among the three $\Gamma_7$ bands are allowed and generates the hybridization gap, which is around 10 meV now.
We note that in contrast to the LDA results, the semiconductor gap obtained by LDA+Gutzwiller is indirect, which is quite consistent with 
the transport measurements\cite{Demsar:2006}.

%Another very important difference between the LDA and LDA+Gutzwiller results is that the semiconductor gap become indirect only
%by LDA+Gutzwiller, which is quite consistent with the transport measurements.

%The probability of the atomic states

\begin{figure}
\includegraphics[width=0.65\textwidth,clip,angle=0]{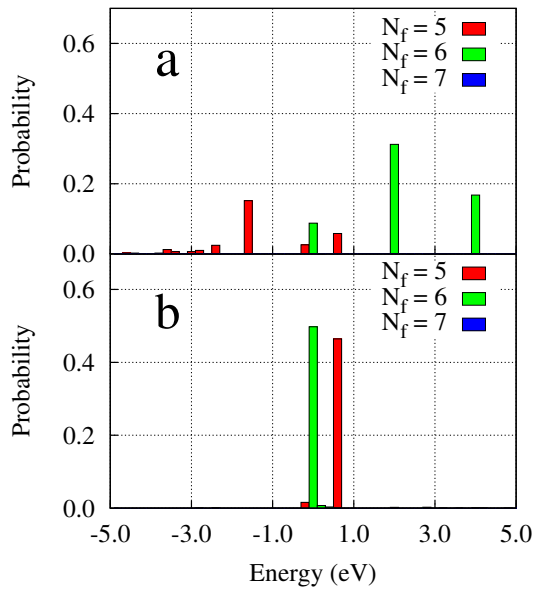}
\caption{The probability of atomic eigenstates in the ground state obtained by (a) LDA alone and (b) LDA+Gutzwiller. $N_f$ is the total number of $f$ electrons for the corresponding atomic eigenstates. The horizontal axis denotes the corresponding atomic eigen energy.  }\label{prob}
\end{figure}

We would emphasize that by LDA+Gutzwiller, we can only obtain the quasi-particle part in the 
spectral function, but not the Hubbard bands containing the atomic multiplet structure\cite{Ylvisaker:2009,Thunstrom:2009}.
While the Gutzwiller type wave function can well capture the multiplet features in the ground state\cite{GW:1963,RTGW:1998,RTGW:2012}. The Gutzwiller variational parameter $\lambda_{\Gamma}$ determines nothing but the probability of each atomic configuration $\Gamma$ in the ground state (which is defined as $\langle G |\Gamma\rangle\langle\Gamma| G\rangle$). In Fig.~\ref{prob}, we plot the corresponding probabilities for SmB$_6$ obtained by LDA+Gutzwiller together with that obtained by LDA wave function($\langle 0 |\Gamma\rangle\langle\Gamma| 0\rangle$).  The LDA ground state is dominated by the atomic configurations with the number of $f$-electrons $N_f$=6, on the other hand however, the distribution of the probability of atomic states obtained by LDA+Gutzwiller is almost equally concentrated on two atomic multiplet states with five and six $f$-electrons respectively, leading to approximately $+2.5$ valence of Sm. 

% inter-mixing of 5/2 and 7/2
Moreover, the Gutzwiller wave function provides
correct description of the intermixing between $j=7/2$ and $j=5/2$ orbitals, which can not be captured by LDA only calculation and manifest itself in the average occupancy of the f orbitals. The occupancy of $j=5/2$ and $j=7/2$ orbitals are $5.31$ and $0.22$ respectively using the LDA type wave function ($|0\rangle$). While the f orbital occupancy is modified to be $3.64$ for $j=5/2$ and $1.89$ for $j=7/2$ orbitals using the Gutzwiller type wave function ($|G\rangle$). The dramatic increment of the  occupancy for $j=7/2$ orbitals is the important consequence of the $F_2-F_6$ terms in the atomic interactions, which can not be expressed by a pure "density-density" form in any single particle basis and generate strong multi-configuration nature for the ground state wave function.

%double couting

% Z2 topological index obtained by LDA+G

\begin{table}[!h]
\caption{The products of parity eigenvalues of the occupied states for TRIM points, $\Gamma$, $X$, $R$ and $M$ in the BZ.}\label{parity}
\begin{tabular*}{0.35\textwidth}{@{\extracolsep{\fill}}cccc}
  \hline\hline
  $\Gamma$ & 3X & R & 3M \\
  \hline
  $+$  & $-$ & $+$ & $+$ \\
  \hline\hline
\end{tabular*}
\end{table}

The band inversion feature around the $X$ points is well persisted in the LDA+Gutzwiller quasi-particle bands. As discussed in previous paragraphs, the topological nature of this interacting system is fully determined by the Gutzwiller effective Hamiltonian $H_{eff}$. Since the spacial inversion symmetry is present for SmB$_6$, we are now able to determine its topological nature by simply counting the parities of those occupied quasi-particle states at 8 TRIM points.  As listed in Table \ref{parity}, the parities are all positive except the X points. Because there are totally three equivalent $X$ points in the whole BZ, the $Z_2$ topological index for SmB$_6$ has to be odd, resulting in a strongly correlated topological insulators with topological indices $(1;111)$.

%The surface band structure obtained bu LDA+G
%Since the band inversion in $SmB_6$ happens at the three equivalent $X$ points, which projects to the $\Gamma$ and two $X$ points 
%of the surface BZ for the $(001)$ surface. 
%are plotted in Fig.\ref{surface} 

\begin{figure}[!h]
\includegraphics[width=0.35\textwidth,angle=90,clip]{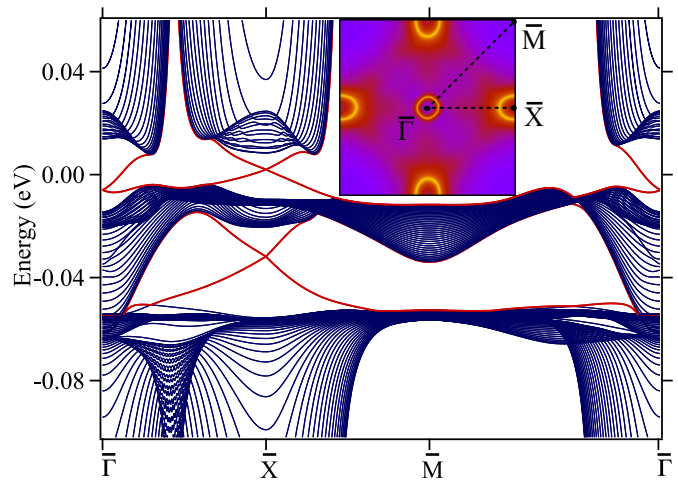}
\caption{The surface states of SmB$_6$ on the (001) surface. The surface states are obtained by LDA+Gutzwiller with a 40 layer slab on the basis of projected Wannier functions. (Inset) The fermi surfaces for SS on (001) surface.
%Three surface Dirac cones are seen at both $\Gamma$ and two $X$ points.
}\label{surface}
\end{figure}

To see the topological surface states, we construct a tight binding model using the projected Wannier functions\cite{Solovyev:2007ce,Kunes:2010p7361,Mostofi:2008ff}, which can reproduce the LDA band dispersion quite precisely.
%{\color{red} (up to 95\% percents near the fermi level?)}.  
The surface states (SS) of the $(001)$ surface are then obtained by combining the above tight binding Hamiltonian and the same rotational invariant Gutzwiller approach on a 40-layers slab. The obtained quasi-particle bands of the slab are plotted in Fig.\ref{surface}. It is clearly seen that the surface states contain three Dirac cones: one cone is located at $\bar{\Gamma}$ and the other two are at two $\bar{X}$ points of the surface BZ. The interesting multi-Dirac-cones behavior is quite unique among the existing TIs and is a natural consequence of the band inversion at the bulk $X$ points, which are projected onto to $\bar{\Gamma}$ and two $\bar{X}$ points of the $(001)$ surface BZ. The multiple Dirac cones on the surface of SmB$_6$ may generate interesting physical phenomenas, such as the unique quasi-particle interference pattern in scanning tunneling microscope (STM), which will be studied in our further publications.

% which reveals three Dirac cone like surface bands on both $\Gamma$ and two $X$ points.

%An accurate low energy model Hamiltonian, which can well catch the crystal-field splitting and the hopping parameters among the active $4f$ and 
%$5d$ orbitals around fermi level in SmB$_6$, is firstly constructed by using projected atomic Wannier functions\cite{mlwf:prb1997, mlwf:prb2001, 
%weng:prb2009} (PAW) method by using OpenMX software package\cite{openMX} within LDA, which can reproduce the calculations of WIEN2k 
%very well. The obtained PAW orbitals are quite localized and the band structure interpolated from the tight-binding (TB) Hamiltonian using these 
%PAW as basis can reproduce that from first-principles calculation. The atomic spin-orbit coupling is added to the above TB Hamiltonian. The SOC
%strength for $4f$ and $5d$, $\eta_{4f}$ and $\eta_{5d}$, are found to be 0.35 and 0.1 eV, respectively, by fitting the band structure from self-
%consistent LDA+SOC calculations.

%conclusion

In summary, we have developed the LDA+Gutzwiller method incorporating the Green's function scheme to study the the topological phases of strongly correlated materials from the first-principles (beyond LDA and LDA+U). This method is systematically applicable to all correlated compounds as long as the quasi-particle weight is not reaching zero. Both quasi-particle bands and atomic multiplete structures can be well captured in the present technique. Applying this method onto typical mixed valence compound SmB$_6$, we demonstrate that it is a strongly correlated 3D TI with unique surface states containing three Dirac cones on the $(001)$ surface. The strong interaction among the $f$ electrons reduces both the band width and the quasiparticle weight for almost one order, but the topological feature remains.

% In summary, using LDA+Gutzwiller we have demonstrated that $SmB_6$ is a 3D strong TI with unique surface band structure containing three Dirac cones in the $(001)$ surface. The strong interaction among the f electrons reduces both the band width and the quasiparticle weight for almost one order, but the topological features keep unchanged.

 This work was supported by the National Science
Foundation of China and by the 973 program of China (No. 2011CBA00108 and 2013CBP21700). We acknowledge the helpful 
discussion with professor P. Coleman, O. Eriksson  and Dr. P. Thunstrom.

%\begin{references}

\bibliography{smb6_v4.0}

%\end{references}

\end{document}